\documentclass[useAMS,usenatbib]{mn2e}
\usepackage{graphicx}
\usepackage{braket}
\usepackage{hyperref}

\title[Distinguishing short and long {\rm Fermi} GRBs]{Distinguishing short and long \textit{Fermi} gamma-ray bursts}
\author[M. Tarnopolski]{M. Tarnopolski\thanks{E-mail:
mariusz.tarnopolski@uj.edu.pl}\\
Astronomical Observatory, Jagiellonian University, Orla 171, 30-244 Krak\'{o}w, Poland}
\begin{document}

\date{\today}

\pagerange{\pageref{firstpage}--\pageref{lastpage}} \pubyear{2015}

\maketitle

\label{firstpage}

\begin{abstract}
Two classes of gamma-ray bursts (GRBs), short and long, have been determined without any doubts, and are usually ascribed to different progenitors, yet these classes overlap for a variety of descriptive parameters. A subsample of 46 long and 22 short {\it Fermi} GRBs with estimated Hurst Exponents (HEs), complemented by minimum variability time-scales (MVTS) and durations ($T_{90}$) is used to perform a supervised Machine Learning (ML) and Monte Carlo (MC) simulation using a Support Vector Machine (SVM) algorithm. It is found that while $T_{90}$ itself performs very well in distinguishing short and long GRBs, the overall success ratio is higher when the training set is complemented by MVTS and HE. These results may allow to introduce a new (non-linear) parameter that might provide less ambiguous classification of GRBs.
\end{abstract}


\section{Introduction}
\label{intro}

Gamma-ray bursts (GRBs) are highly energetic events that often are brighter than all other $\gamma$-ray objects visible on the sky combined, with an emission peak in the 200--500 keV region \citep[for recent reviews, see][]{nakar,zhang4,gehrels2,berger,meszaros}. GRBs were detected by military satellites \textit{Vela} in late 1960's. \citet{mazets} first observed a bimodal distribution of $T_{90}$ (time during which 90\% of the burst's fluence is accumulated) drawn for 143 events detected in the KONUS experiment. BATSE onboard the Compton Gamma Ray Observatory ({\it CGRO}) \citep{meegan92} allowed to confirm the hypothesis \citep{klebesadel} that GRBs are of extragalactic origin due to isotropic angular distribution in the sky. However, a more complete sample of BATSE short GRBs were shown to be distributed anisotropically and cosmological consequences were discussed lately by \citet{meszaros}. BATSE 1B data release was followed by further investigation of the $T_{90}$ distribution \citep{kouve} that lead to establishing the common classification of GRBs into short ($T_{90}<2\,{\rm s}$) and long ($T_{90}>2\,{\rm s}$), and based on which GRBs are most commonly classified. The progenitors of long GRBs are associated with supernovae \citep{woosley} related with collapse of massive stars. Progenitors of short GRBs are thought to be NS-NS or NS-BH mergers \citep{nakar}, and no connection between short GRBs and supernovae has been proven \citep{zhang5}.

Despite initial isotropy, short GRBs were shown to be distributed anisotropically on the sky, while long GRBs are distributed isotropically \citep{balazs,maglio,vavrek} (see also \citealt{meszaros2} for an analysis that revealed anisotropy in a set of long GRBs). Since long a possibility that GRBs may be divided into more than two classes was put forward as there are GRBs that do not fall easily into one class or another \citep{gehrels,zhang,bromberg}. \citet{horvath98,horvath02} investigated GRBs from the BATSE catalog \citep{meegan96,paciesas} by a univariate approach and concluded that the probability that the peak at intermediate values of $\log T_{90}$ is a result of random fluctuations is much less than $1\%$. Also in {\it Swift} data evidence for a third component in $\log T_{90}$ was found \citep{horvath08a,horvath08b,zhang,huja,huja2,horvath10}. Other datasets, i.e. {\it RHESSI} \citep{ripa,ripa2}, or {\it BeppoSAX} \citep{horvath09}, also show evidence for an intermediate class. It is worth to note that the intermediate GRBs are distributed anisotropically on the sky \citep{meszaros3,meszaros4,vavrek}. The origin and existence of an intermediate class remains elusive as theoretical models still need to account for an apparent bimodality in duration distribution \citep{janiuk,nakar}. Finally, not only the existence of an intermediate class was investigated (and remains unsettled), but also subclass classifications of long GRBs were proposed \citep{gao}.

The research conducted on {\it Fermi} data \citep{gruber,kienlin} are consistent with a bimodal duration distribution \citep{zhang3} as well as with the existence of an intermediate class \citep{horvath12}; however, even a bimodal structure was not present in some energy bands in the examined sample of 315 GRBs (not to mention trimodality) and may be due to an instrumental selection effect \citep{qin}. It is important to note that in all mentioned research only a mixture of standard normal distributions (Gaussians) was fitted to the observed $\log T_{90}$ distributions. By examining the duration distribution it was shown that the third class is unlikely to be present in the {\it Fermi} data \citep{Tarnopolski}, and that a mixture of two intrinsically skewed distributions follows better the $\log T_{90}$ distribution of {\it Fermi} GRBs than a mixture of three standard Gaussians \citep{Tarnopolski3}.

The {\it Fermi} dataset has been examined widely for other reasons, too. Its redshift distribution was investigated \citep{ackermann} confirming the observation that short GRBs have lower redshifts than long GRBs. The Amati correlation was investigated \citep{basak,gruber2} and a link between short and long GRBs was discovered \citep{muccino}. It gave insight into the GRB afterglow population \citep{racusin}, allowed to observe a number of high-energy GRBs (with photon energies exceeding 100~MeV, or even 10~GeV) \citep{atwood}, and provided a verification of the short--long classification \citep{zhang3}.

Other parameters, despite $T_{90}$ durations (e.g., similarly defined $T_{50}$ or hardness ratios for various bands) were used to classify GRBs for different satellite databases as well. A distinction in the hardness ratio $H_{32}$ of all three classes was formulated early \citep{mukh,horvath04}. Minimum variability time-scale (MVTS) \citep{bhat,maclach12,maclach13a,maclach13b} was used as a distinctive parameter for differentiating GRB lightcurves based on their temporal structure. It is a non-parametric feature, such as the Hurst exponent (HE, denoted $H$) \citep{hurst,mandel68}, which is a quantitative measure of the persistence of the signal, i.e. it reveals a temporal trend in the overall behavior in the data, and was suggested to be applicable in GRB classifications \citep{maclach13b}.

Among many existing computational algorithms for estimating $H$ (Rescaled Range Analysis \citep{mandel69}, Detrended Fluctuation Analysis \citep{peng94,peng95}, wavelet approach \citep{simons}, Detrended Moving Average \citep{carbone}, to mention only a few), their extraction for the {\it Fermi} GRB sample was performed by the wavelet approach with the Haar wavelet as a basis (for further details, see \citealt{maclach13b} and Sect.~\ref{data2} herein for a brief description). It was found that there is an offset between means of $H$ distributions of long and short GRBs that may serve as a criterion for distinguishing these two classes. Moreover, it was proposed that the overlap region of HE distribution may be related to the third class of GRBs (possibly intermediate). However, the region where short and long GRBs overlap is significant due to a large dispersion in $H$, and the union distribution is skewed toward values less than $0.5$. 

Additional parameters have been defined and proposed for GRB classification as well. Examples are \mbox{$\varepsilon=E_{\gamma,{\rm iso},52}/E_{p,z,2}^{5/3}$} (where $E_{\gamma,{\rm iso},52}$ is the isotropic gamma-ray energy given in $10^{52}\,{\rm erg}$, and $E_{p,z,2}$ is the cosmic rest-frame spectral peak energy in units of $100\,{\rm keV}$) unambiguously dividing short and long GRBs \citep{lu}, MVTS \citep{bhat,maclach12,maclach13a,maclach13b,golkhou,golkhou2} or HE \citep{maclach13b}. However, providing a set of parameters that could classify GRBs with high accuracy remains a desired objective. Unsupervised machine learning (ML) algorithms, trained on a dataset containing these parameters, yield a possibility of attaining this goal. Before that, performance of supervised ML with arbitrary number of classes needs to be tested.

In this work, focus is laid on performance of HEs, complemented by MVTS (both being wavelet-based parameters new in GRB classification) and $T_{90}$, being a classical GRB type indicator, in classifying short and long GRBs. This paper is organized in the following manner. In Sect.~\ref{data} a description of the data is provided. In Sect.~\ref{HEGRBs} a statistical analysis of the available dataset of GRBs with computed HEs is conducted. In Sect.~\ref{ML}, ML is applied to $(H,MVTS,T_{90})$ singles, pairs and triples, and a Monte Carlo simulation provides statistics of the success ratio. Section~\ref{discussion} contains discussion of the results, and Sect.~\ref{conclusions} gives concluding remarks. A computer algebra system {\sc Mathematica}\textsuperscript{\textregistered} v10.0.2 is used throughout this paper.

\section{The sample}
\label{data}

\subsection{Selection criteria}
\label{data0}

The data are taken from \citet{maclach13a,maclach13b}, and are complemented by \citet{golkhou,golkhou2}. They consist of 46 long and 22 short GRBs detected by the {\it Fermi} satellite. According to \citet{maclach13a}, the selection criteria for the GRB sample examined in \citet{maclach13b} are two-fold. The following condition on the ratio of variances for one or more octaves $j$,
\begin{equation}
\frac{\beta_j^{\rm preburst}}{\beta_j^{\rm burst}}<0.75,
\label{eq1}
\end{equation}
was required, where $\beta_j^{\rm preburst}$ corresponds to a certain time interval before the burst\footnote{For long GBs the preburst was defined to start $20\,{\rm s}$ and finish $5\,{\rm s}$ before the trigger, and for short GRBs these were $15\,{\rm s}$ and $1\,{\rm s}$, respectively.}, used as a surrogate for the background variance, and $\beta_j^{\rm burst}$ expresses the variability of the burst. The variances were computed according to
\begin{equation}
\beta_j=\frac{1}{n_j}\sum\limits_{k=0}^{n_j-1}|d_{j,k}|^2,
\label{eq2}
\end{equation}
where $d_{j,k}$ are the wavelet transform coefficient (for further details, see \citealt{maclach13b} and Sect.~\ref{data2} herein for a brief explanation). Additionally, it was required that the lightcurve fits had a $\chi_{\rm reduced}^2<2$.

\subsection{Duration $T_{90}$}
\label{data1}

The duration $T_{90}$ is the time during which the cumulative counts increase from 5\% to 95\% above background, encompassing 90\% of the total fluence detected. This way of measuring the duration is independent of the intensity for a given instrument \citep{kouve}. The detection is triggered when the signal exceeds a certain threshold.

{\it Fermi} has a time resolution of $2\mu{\rm s}$; the lightcurves are extracted at a binning of $200\mu{\rm s}$. Time intervals that are far before and far after the main burst are selected as background and subtracted from the lightcurve. The procedure is described in detail by \citet{koshut}.

\begin{figure}
\begin{center}
\includegraphics[width=0.99\linewidth]{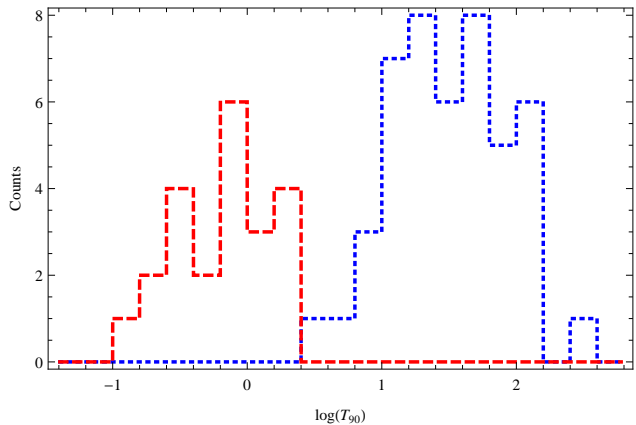}
\end{center}
\caption{Distributions of $\log T_{90}$: dashed red -- short GRBs, dotted blue -- long GRBs.}
\label{fig1}
\end{figure}

The durations used herein come from the {\it Fermi} GRB Burst-Catalogue as cited in \citet{maclach13b}. They are also available in the online catalogue\footnote{\url{http://heasarc.gsfc.nasa.gov/W3Browse/fermi/fermigbrst.html}}. The $T_{90}$ distribution of GRBs examined herein is displayed in Fig.~\ref{fig1}.

\subsection{Hurst Exponent}
\label{data2}

The HE describes the level of statistical self-similarity of a time-series. A time-series $X(t)$ with $N$ data points $t_i$ is called self-similar (or self-affine) with a Hurst exponent $H$ if, after a rescaling $t\rightarrow\lambda t$, it satisfies the following relation
\begin{equation}
X(t)\stackrel{\textbf{\textrm{.}}}{=}\lambda^{-H}X(\lambda t),
\label{eqA}
\end{equation}
where $\stackrel{\textbf{\textrm{.}}}{=}$ denotes equality in distribution. The HEs under study were calculated by \citet{maclach13b} with the fast wavelet transform using the Haar wavelet as the basis. The basis is obtained from a mother wavelet, $\psi (t)$, by
\begin{equation}
\psi(t)\rightarrow\psi_{j,k}(t)=\frac{1}{2^{j/2}}\psi\left(\frac{t-k}{2^j}\right),
\label{eqB}
\end{equation}
where $j$ represents the octave (time-scale) and $k$ the position of the wavelet. The wavelet transform coefficient is defined as
\begin{equation}
d_{j,k}=\braket{X,\psi_{j,k}}.
\label{eqC}
\end{equation}

By inserting Eq.~\ref{eqA} into Eq.~\ref{eqC} one gets the relation between the variance of the wavelet coefficients $d_{j,k}$ and the scale $j$:
\begin{equation}
\log_2 {\rm var}(d_{j,k})=(2H+1)j+{\rm const.}
\label{eqD}
\end{equation}
The HE is obtained by fitting a line to the linear part of the $\log_2 {\rm var}(d_{j,k})$ vs. $j$ relation. For further details, the reader is referred to \citet{maclach13b}.

The HE is bounded in the interval $[0,1]$ and is equal to 0.5 for a random walk (Brownion motion). For persistent data $H>0.5$, while for anti-persistent $H<0.5$. Regular (periodic or quasi-periodic) time series posses $H=1$. It is related to the fractal dimension of the signal $D=2-H$, providing information about its statistical self-similarity. This means that in a persistent process the increments are persistent themselves \citep{clegg} with the same $H$. A Brownian motion has independent increments.

\begin{figure}
\begin{center}
\includegraphics[width=0.99\linewidth]{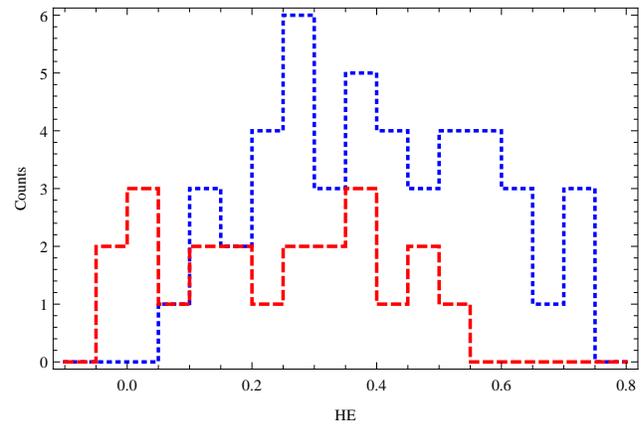}
\end{center}
\caption{Distributions of HEs: dashed red -- short GRBs, dotted blue -- long GRBs.}
\label{fig2}
\end{figure}

The HE distribution of GRBs examined herein is displayed in Fig.~\ref{fig2}. Note there are two small negative values. The negative sign most likely comes about because of poor quality of the lightcurves, and the features of the lightcurves may prevent the algorithm used to converge to a meaningful value (Sonbas, private communication). Most likely these HEs are very small, but positive. However, for completeness reasons and in order not to further diminish the GRB sample used, those values are not rejected from the analysis as it might happen in future research that other lightcurves may also exhibit a negative value in the frame of the method applied (in this case, the wavelet approach). Keeping those two values takes into account numerical artifacts that might be encountered in other GRBs. Nevertheless, as those HEs correspond to short GRBs and are consistently placed in the HE histogram, they should not affect the short/long classification.

\subsection{Minimum Variability Time-Scale}
\label{data3}

If the lightcurve is binned into too narrow bins, the intrinsic variability is lost in the statistical noise. When the binning is too coarse, the variability might disappear. Hence, the right binning is crucial and varies with the lightcurve. To infer the right binning, a comparison of the variances of the GRB and the background is performed. A ratio of these variances is plotted against the bin-width and the MVTS is the binning at which this ratio is minimized \citep{bhat}.

Expressing this concept in other words, MVTS is a time-scale at which a transition between white noise and a power law is observed in the power density spectrum. This means that MVTS marks the time-scale at which a power law process and a white noise exchange dominance. As wavelets are sensitive to whichever process dominates at a given time-scale, they are a natural choice to extract the MVTS from the lightcurves.

The wavelet-based MVTS extraction is carried out similarly to the HE estimation described in Sect.~\ref{data2}. Starting from Eq.~\ref{eqD}, with the background subtraction as described in Sect.~\ref{data1}, the MVTS is calculated at the octave $j_{\rm intersection}$ at wchich an intersection of a flat noise spectrum and a linear relation associated with the power law domain in the log-scale diagram occurs:
\begin{equation}
MVTS=T_{\rm bin}\times 2^{j_{\rm intersection}},
\label{eqE}
\end{equation} 
where $T_{\rm bin}$ is the finest binning of the data. For further details, the reader is referred to \citet{maclach13a}.

\begin{figure}
\begin{center}
\includegraphics[width=0.99\linewidth]{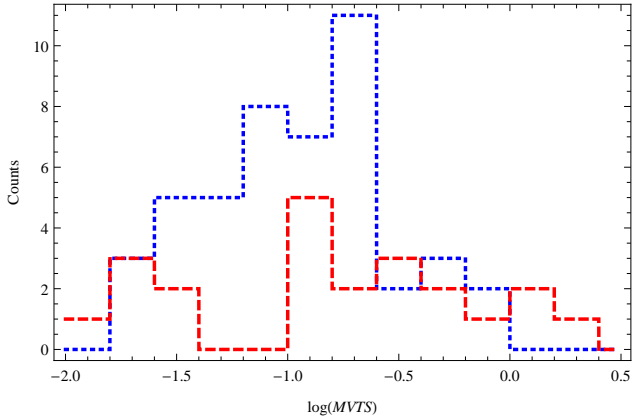}
\end{center}
\caption{Distributions of $\log MVTS$: dashed red -- short GRBs, dotted blue -- long GRBs.}
\label{fig3}
\end{figure}

The MVTS used in this study are taken from \citet{maclach13a}, although some values for the short GRBs subsample were not reported there. Hence, the missing values were replaced with the MVTS from \citet{golkhou,golkhou2}. The distribution of the sample under consideration is shown in Fig.~\ref{fig3}.

\section{Independence of short and long GRBs' HE distributions}
\label{HEGRBs}

In Fig.~\ref{fig2}, the HE distributions for short and long GRBs from \citet{maclach13b} are reproduced. While the systematic difference is clearly visible, the histograms overlap strongly. This leads to a question whether they are statistically distinguishable. As most of statistical tests \citep{kendall,kanji,recipes} assume normality, first a Kolmogorov--Smirnov test \citep{kolmogorov} is performed, that resulted in $p$-values of $0.95$ and $0.87$ for short and long GRBs, respectively. These values are large, so might lead to a conclusion that both samples are normally distributed. However, in the sample of long GRBs, ties are present, and the $p$-value is obtained after ignoring them. A mean of hundred Monte Carlo realisations is equal to $0.8735\pm 0.0102$, where the error is the standard deviation of the mean. Another trial gave $0.8728\pm 0.0106$. To be sure these data are normally distributed two more tests are performed. The Pearson $\chi^2$ test \citep{voinov} gave $p$-values equal to $0.75$ and $0.81$, and the Shapiro--Wilk test \citep{shapiro} gave $0.33$ and $0.39$ for short and long GRBs, respectively. Despite these values span a quite broad interval, they are all significant enough to ascertain that the HE distributions are normal.

To check whether they are separate, first a $t$-test \citep{student} is performed, based on which at the $0.01$ significance level the hypothesis that the means' difference is $\mu_{\rm long}-\mu_{\rm short}=0.163215$ cannot be rejected. The means were computed from the samples. Also, the Mann--Whitney test \citep{mann} does not allow to reject the hypothesis that the medians' difference is $m_{\rm long}-m_{\rm short}=0.144$ at the 0.01 significance level. Finally, variance tests: Brown--Forsythe \citep{brown}, Fisher Ratio \citep{recipes}, and Levene \citep{levene}, all gave $p$-values greater than $0.9$. All of these tests imply that the HE distributions indeed are different for short and long GRBs.


\section{Machine Learning}
\label{ML}

As other schemes of classifying GRBs were proposed, e.g. based on MVTS, it is desired to investigate their efficiency for distinguishing short and long GRBs. As the amount of GRBs classified in such a way is still limited, in the following section focus is paid only to the two classical classes.

\subsection{Methods}
\label{methods}

In order to verify whether $H$ can serve as a GRB type indicator \citep{maclach13b}, a supervised machine learning (ML) is performed. A support vector machine (SVM) algrithm is employed. It aims at building a potentially non-linear model that is trained over a set of examples that are classified as belonging to one of arbitrarily specified categories ({\it short} or {\it long} in this case). The training set is represented in a multi-dimensional space, and the classification is done so that the categories are divided by a gap as wide as possible. After the SVM is trained, a classification probability is derived over the set of applicable parameters. Based on this classification, new examples are mapped into the same space and predicted to belong to one category or another based on which side of the dividing hyperplane they fall on. The probability threshold is set at $50\%$, and no undetermined classifications are allowed.

The parameter space is three-dimensional, and constitutes of $H$, MVTS and $T_{90}$. It is found that the number of mistaken classifications is reduced when using a space of $H$, $\log MVTS$ and $\log T_{90}$, and this semi-log space is explored throughout this section. A subsample of the total 22~short and 46~long GRBs is taken to be a training set, while the remainings GRBs serve as a validation set. A subsample is created in the following way. For each training set, 20 out of 22 short GRBs are chosen. The number of these subsets is equal to ${22 \choose 20} = 231$. Among long GRBs, 42 out of 46 examples are randomly chosen for each of the short GRBs subsets, and this is repeated 435 times. A total of $\sim 10^5$ SVM models is built, and they exhaust the number of realisations of short GRBs subsets\footnote{There is a total of ${22 \choose 20} \times {46 \choose 42} \approx 3.77 \cdot 10^7$ subsets.}.

ML is done in seven spaces:
\begin{enumerate}
\item three univariate classifications based on $H$, $\log MVTS$ and $\log T_{90}$,
\item three bivariate classifications based on pairs $(H,\log MVTS)$, $(H,\log T_{90})$ and $(\log T_{90},\log MVTS)$,
\item a trivariate classification based on triples $(H,\log MVTS,\log T_{90})$.
\end{enumerate}
For each, a success ratio $r$ is inferred from the number of correct classifications in the following manner. If both short GRBs from the validation set (consisting of two short and four long GRBs) were classified correctly by an SVM model, the success ratio $r_{\rm short}$ is equal to 1; if one is correct and the other is false, $r_{\rm short}=0.5$; and if both were classified falsely as long GRBs, $r_{\rm short}=0$. Similarly for long GRBs, $r_{\rm long}$ is computed, and takes the values 0, 0.25, 0.5, 0.75 or 1. A~total success ratio is defined as
\begin{equation}
r_{\rm tot}=\frac{2r_{\rm short}+4r_{\rm long}}{6},
\label{eqF}
\end{equation}
and the success ratios are calculated for all $\sim 10^5$ trials. The relative frequency for each $r$ is obtained by dividing the counts by the number of trials.

\subsection{Results}
\label{results}

\subsubsection{1D parameter spaces}
\label{1d}

\begin{figure*}
\begin{center}
\includegraphics[width=0.99\linewidth]{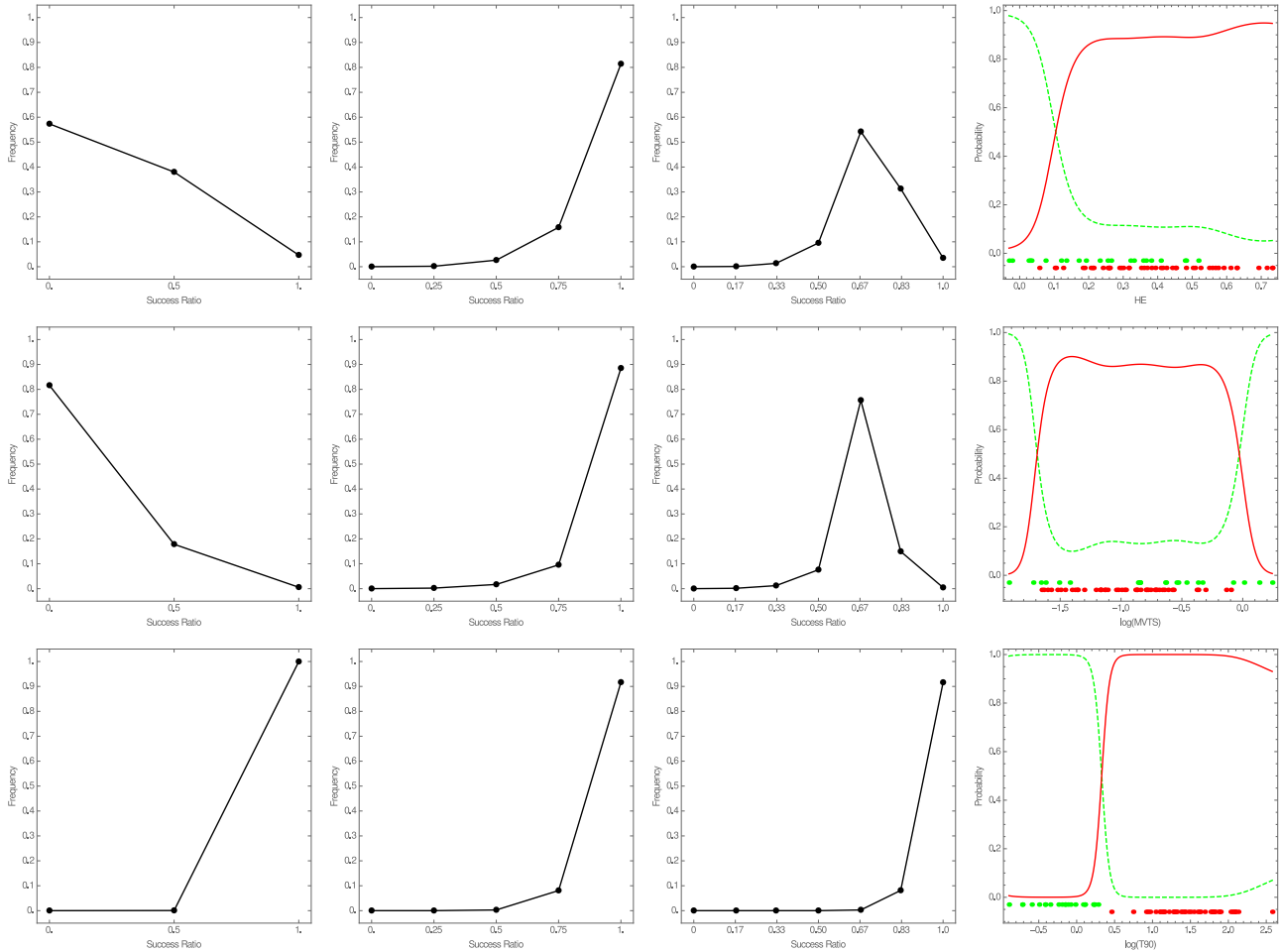}
\end{center}
\caption{Frequencies of successful univariate predictions. {\it Rows}, from upper: based on HEs, on MVTS, and on durations $T_{90}$. {\it Columns}, from left: fraction of successes for short GRBs, for long GRB, and for a union of both. Right panel displays examples of probability functions learnt by SVM. Dashed green -- short GRBs, solid red -- long GRBs. Dots mark the training set: green (upper row) -- short GRBs, red (bottom row) -- long GRBs.}
\label{fig4}
\end{figure*}

The results of ML for univariate distributions of parameters under consideration are displayed in Fig.~\ref{fig4}. The HEs, suggested to be a candidate for distinguishing between short and long GRBs \citep{maclach13b}, exhibit a rather poor performance: $57\%$ of short GRBs were classified improperly, and correct matches took place for only $5\%$ of them. On the other hand, $81\%$ of long GRBs were classified correctly. This leads to a peak in total matches at the level of $54\%$ for the success ratio $r_{\rm tot}=0.67$. Overall, this is an unsatisfactory result that comes from a large overlap of HEs for short and long GRBs (compare with Fig.~\ref{fig2}).

Using $\log MVTS$ similar results are obtained, with an even higher frequency of mismatches for short GRBs, equal to $82\%$. On the other hand, it performs slightly better in distinguishing long GRBs with an $r_{\rm long}=1$ in $88\%$ of validation sets. This leads to a peak of $76\%$ at $r_{\rm tot}=0.67$, which is a considerable accuracy.

Historically, the distinction between short and long GRBs was made based on $\log T_{90}$ distribution \citep{kouve}, and the classification is straightforward: if $T_{90}<2\, {\rm s}$ ($T_{90}>2\, {\rm s}$), a GRB is short (long). Nevertheless, given that SVM classification is probability-based, one can expect a lower than $100\%$ accuracy, especially that both short and long GRBs may be arbitrarily close to the limit $T_{90}=2\,{\rm s}$. Indeed, short (long) GRBs were properly recognized in $99.95\%$ ($91.64\%$) cases, what gives a net accuracy of $91.59\%$ (at $r_{\rm tot}=1$). However, as $\log T_{90}$ ML is effective for short GRBs, while $\log MVTS$ and $H$ for long ones, their combination is likely to give more reliable results. For both HE and $\log MVTS$ distributions, the probabilities favor long GRBs (compare with right column in Fig.~\ref{fig4}).

\subsubsection{2D parameter spaces}
\label{2d}

\begin{figure*}
\begin{center}
\includegraphics[width=0.99\linewidth]{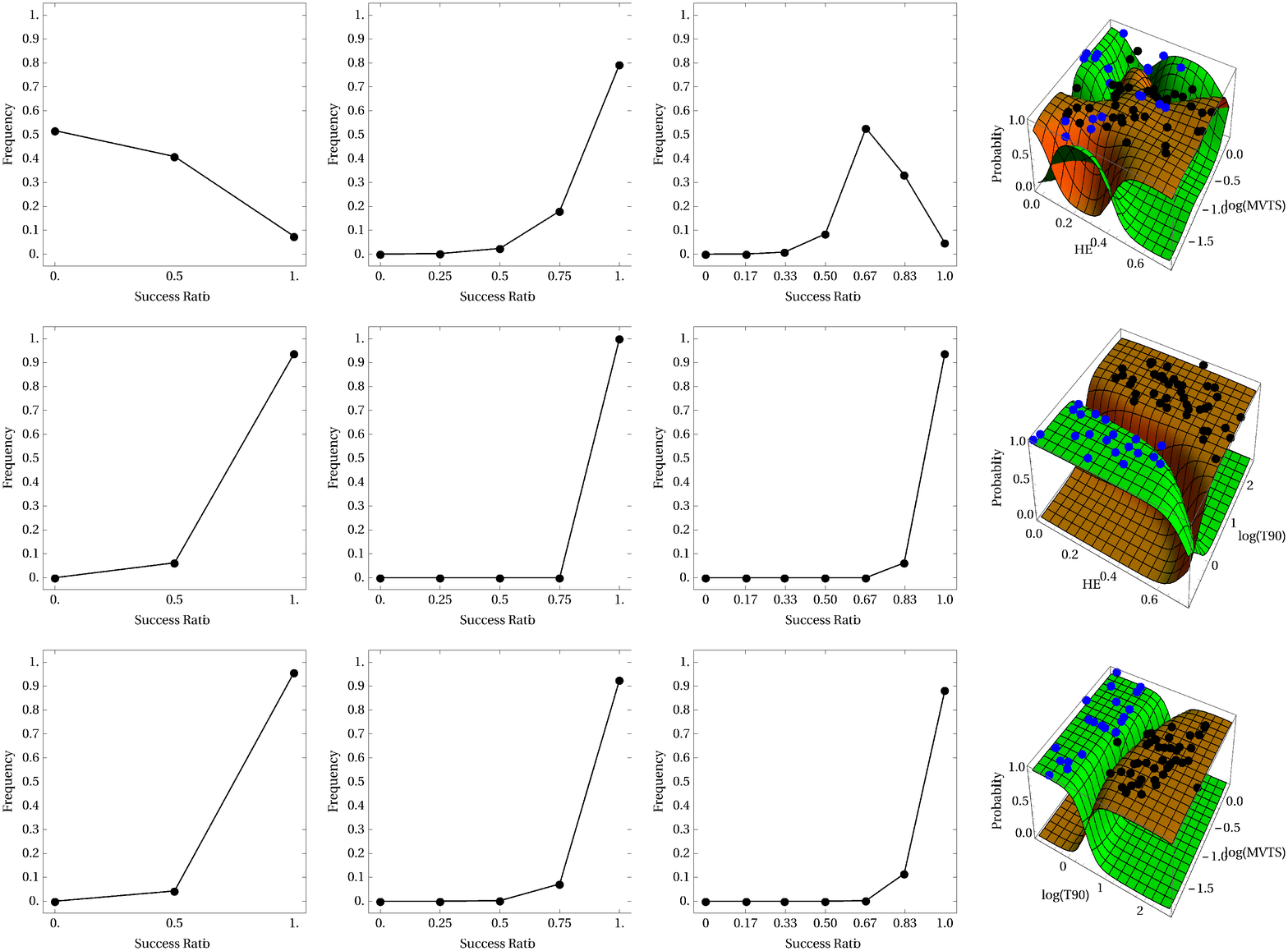}
\end{center}
\caption{Frequencies of successful bivariate predictions. {\it Rows}, from upper: based on HEs and MVTS, on HEs and durations $T_{90}$, and on MVTS and durations $T_{90}$. {\it Columns}, from left: successes for short GRBs, for long GRB, and for a union of both. Right panel displays examples of probability functions learnt by SVM. Green surface -- short GRBs, red surface -- long GRBs. Dots mark the training set: blue -- short GRBs, black -- long GRBs.}
\label{fig5}
\end{figure*}

As both $H$ and $\log MVTS$ distributions exhibit large overlaps for GRB populations, the results of ML displayed in Fig.~\ref{fig5} for a set of pairs $(H,\log MVTS)$ are neither surprising, nor interesting. Short GRBs are falsely classified in $52\%$ of cases, while long GRBs are correctly recognized in $79\%$ of validation sets. This gives a peak of $53\%$ at $r_{\rm tot}=0.67$.

However, when HEs are complemented with $\log T_{90}$, an accuracy of $100\%$ for long GRBs is attained. For short GRBs, an accuracy of $94\%$ is obtained, what results in a~total proper recognition level of also $94\%$. Note this is a~better result than for $H$ and $\log T_{90}$ alone.

Distributions in $\log MVTS$ have larger overlaps than in $H$, hence the performance of $\log T_{90}$ complemented with MVTS is slightly worse than in the previous case: success ratios equal to unity are obtained in $96\%$ and $93\%$ for short and long GRBs, recpectively, and $r_{\rm tot}=1$ in $88\%$ cases. Considering this space as durations complemented with MVTS, the accuracy is lower. On the other hand, when MVTS are complemented with $\log T_{90}$, the performance is much better than for MVTS alone. Again, when the populations overlap strongly, the probabilities favor long GRBs (compare with right column in Fig.~\ref{fig5}).

\subsubsection{3D parameter spaces}
\label{3d}

\begin{figure*}
\begin{center}
\includegraphics[width=0.99\linewidth]{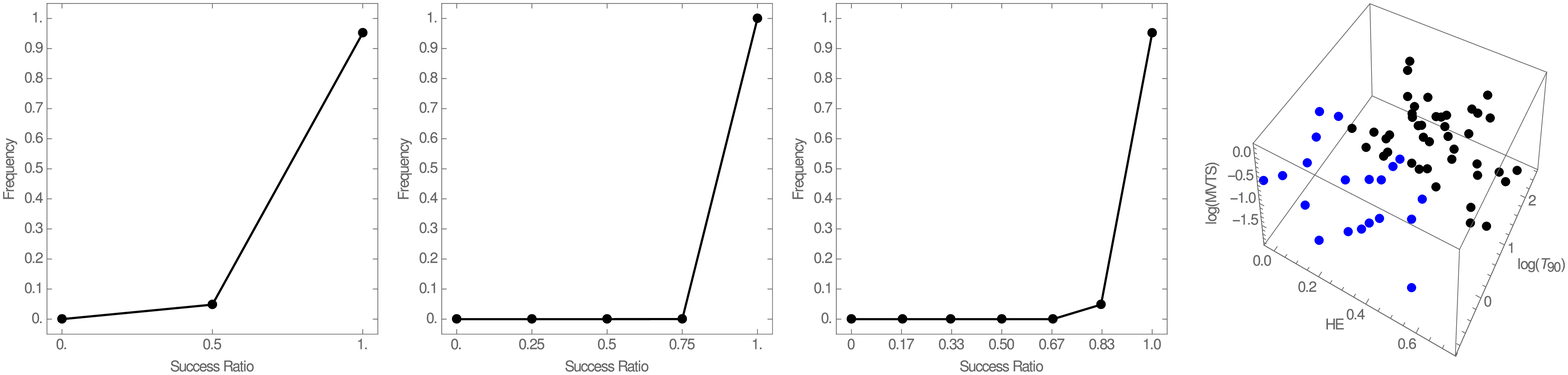}
\end{center}
\caption{Frequencies of successful trivariate predictions. From left: short GRBs, long GRBs, and a union of both. Right panel displays an example training set for a trivariate machine learning. Blue dots -- short GRBs, black dots -- long GRBs.}
\label{fig6}
\end{figure*}

When ML is performed on a three-dimensional training set (see right panel in Fig.~\ref{fig6}) of triples $(H,\log MVTS,\log T_{90})$, short GRBs are classified properly in $95.18\%$, long GRBs in $99.99\%$, and the overall accuracy is higher than $95\%$. Comparing with 1D and 2D approaches, this gives a significant improvement (see Fig.~\ref{fig6}). Specifically, when $(\log T_{90},\log MVTS)$ is complemented with HEs, the gain in accuracy is almost $7\%$. It is not surprising, however, that complementing $(H,\log MVTS)$ with durations gives a significant improvement, as $T_{90}$ allows to introduce a clear separation between the populations, which are highly mixed otherwise. Moreover, adding $\log MVTS$ to $(H,\log T_{90})$ gives an improvement of about $1.5\%$.

\section{Discussion}
\label{discussion}

Unsupervised ML gives hints about the number of classes in a dataset. This is, however, a phenomenological approach that does not have to reflect differences in physical processes underlying their origin \citep{horvath06,nakar,bromberg}. Given the small dataset with computed HEs, a supervised ML was performed in order to verify whether the features recently introduced to characterize GRBs, i.e., MVTS \citep{bhat,maclach12,maclach13a} and HE \citep{maclach13b}, are suitable for automated classification. As the commonly accepted division of GRBs into short and long according to $T_{90}$ is its key, based on which the classes for an SVM classifier were attributed to available MVTS and HE values, a three-dimensional parameter space $(H,\log MVTS,\log T_{90})$ was studied. It was shown that the accuracy was much greater in this parameter space than in a one- or two-dimensional parameter spaces. Especially, complementing $(\log MVTS,\log T_{90})$ by $H$ lead to a better performance. Therefore, the HE is considered a candidate indicator for distinguishing short and long GRBs, however its performance alone was rather poor. On the other hand, a sample of $22+46=68$ GRBs with calculated HEs is not numerous enough to adjudicate with certainty what is the genuine overlap for the whole population. 

Another issue is to construct the training set; specifically, to unambiguously attribute classes to the objects observed. The conservative limit of $T_{90}=2\,{\rm s}$ was shown to be appropriate for BATSE GRBs, but for {\it Swift} data a division at $T_{90}=0.8\,{\rm s}$ is more suitable \citep{bromberg}. The plausible limits were derived (by means of a local minimum of a bimodal distribution of $T_{90}$ durations) for a number of catalogs by \citet{Tarnopolski2} and gave values in the range $2.05\,{\rm s}$ to $3.38\,{\rm s}$, hence the classification is detector dependent\footnote{It appears that the {\it Fermi} dataset has a limit of $2.05\,{\rm s}$ \citep{Tarnopolski2}, very close to the conventional limit derived by \citet{kouve}.}. Note that the short and long associations for the current study were taken from \citet{maclach13a,maclach13b,golkhou,golkhou2}, and are based on the classical limit of $2\,{\rm s}$. Moreover, GRBs classified as short often are of collapsar origin and those classified as long are recognized as having non-collapsar progenitors \citep{nakar,bromberg}. Other properties commonly used as indicators are the hardness ratio, spectral lag, peak energy etc. \citep{bagoly,borgonovo,zhang5}, but they are still selected phenomenologically. It was also suggested that GRBs may be generated by a third mechanism, most likely shock break out \citep{bromberg2}.

It was shown in a number of research, for BATSE data \citep{horvath06} as well as for {\it Swift} sample \citep{veres,horvath10}, that while the GRB separation by duration or hardness ratio alone is ambiguous, the joint distribution in a two-dimensional space of $\log H_{32}$ vs. $\log T_{90}$ allows a sharper differentiation between the GRB types. In the spirit of these findings, and in light of the proposition that HEs may serve as a GRB type indicator \citep{maclach13b}, it was verified that complementing the sets of $\log T_{90}$ and $(\log T_{90},\log MVTS)$ by HEs increases the success rate defined in Eq.~(\ref{eqF}). The fact that HE and MVTS alone, i.e. without $T_{90}$, are less convenient is not fully new, but the main result herein is that HEs help in distinguishing GRBs of different types.

SVM performs well in distinguishing short and long GRBs with data complemented by HEs. After setting the number of physically different GRB classes (at least two, and possibly three), which will require a working model that may hopefully predict the distributions  (univariate, i.e. durations, hardness ratio, etc., as well as multivariate) of characteristic features, a supervised ML might then be applied to automated classification of GRBs observed in the future.

\section{Conclusions}
\label{conclusions}

In this paper, GRBs observed by {\it Fermi} were investigated. Among $\sim 1600$ GRBs, there are 68 GRBs with calculated HE values, and a supervised ML was performed on this subsample. The SVM method is applied to singles, pairs and triples of $(H,\log MVTS,\log T_{90})$. The HE and MVTS (parameters new in GRB classification) alone perform rather poor in distinguishing between short and long GRBs due to a large overlap in their distributions. However, when the pairs $(\log MVTS,\log T_{90})$ are complemented by HEs, the accuracy of classification increases by 7\% and exceeds 95\%, i.e., introduction of $H$ in the classification scheme gives more accurate results. Hence, HEs are suggested as indicator candidates for distinguishing short and long GRBs, however not alone, as it is commonly done with durations $T_{90}$, but as part of an association of parameters. Particularly, in this paper an association of three parameters $(H,MVTS,T_{90})$ was examined.

It is suggested that HEs will be useful in classifying GRBs after the number of their classes is unambiguously determined, possibly by constructing a working model rather than by a phenomenological approach. As the shape of the HE distribution could not be determined accurately due to a small sample (consisting of 68 GRBs), a bigger, preferably complete, sample of {\it Fermi} GRBs might reveal new properties of the GRB population that could either challenge present theoretical models, or give hints on how to construct them.

\bsp

\label{lastpage}


\begin{thebibliography}{99}

\bibitem[\protect\citeauthoryear{Ackermann et al.}{2013}]{ackermann} Ackermann M., 2013, ApJS, 209, 11
\bibitem[\protect\citeauthoryear{Atwood et al.}{2013}]{atwood} Atwood W. B. et al., 2013, ApJ, 774, 76
\bibitem[\protect\citeauthoryear{Allessio et al.}{2002}]{carbone} Alessio, E., Carbone, A., Castelli, G., Frappietro, V., 2002, Eur. Phys. J. B, 27, 197
\bibitem[\protect\citeauthoryear{Bagoly et al.}{1998}]{bagoly} Bagoly Z., M\'esz\'aros A., Horv\'ath I., Bal\'azs L. G., M\'esz\'aros P., 1998, ApJ, 498, 342
\bibitem[\protect\citeauthoryear{Bal\'azs, M\'esz\'aros \& Horv\'ath}{1998}]{balazs} Bal\'azs L. G., M\'esz\'aros A., Horv\'ath I., 1998, A\&A, 339, 1
\bibitem[\protect\citeauthoryear{Basak \& Rao}{2013}]{basak} Basak R., Rao A. R., 2013, MNRAS, 436, 3082
\bibitem[\protect\citeauthoryear{Berger}{2014}]{berger} Berger E., 2014, ARA\&A, 52, 43
\bibitem[\protect\citeauthoryear{Bhat}{2013}]{bhat} Bhat P. N., 2013, in Castro-Tirado A. J., Gorosabel J., Park I. H., eds,
EAS Publ. Ser. Vol. 61, Temporal Decomposition Studies of GRB Lightcurves. Cambridge Univ. Press, Cambridge, p. 45
\bibitem[\protect\citeauthoryear{Borgonovo \& Bj\"ornsson}{2006}]{borgonovo} Borgonovo L., Bj\"ornsson C.-I., 2006, ApJ, 652, 1423
\bibitem[\protect\citeauthoryear{Bromberg, Nakar \& Piran}{2011}]{bromberg2} Bromberg O., Nakar E., Piran T., 2011, ApJ, 739, L55
\bibitem[\protect\citeauthoryear{Bromberg et al.}{2013}]{bromberg} Bromberg O., Nakar E., Piran T., Sari R., 2013, ApJ, 764, 179
\bibitem[\protect\citeauthoryear{Brown \& Forsythe}{1974}]{brown} Brown M. B., Forsythe A. B., 1974, J. Amer. Statist. Assoc., 69, 364
\bibitem[\protect\citeauthoryear{Clegg}{2006}]{clegg} Clegg R. G., 2006, Int. J. Simulation, 7, 3
\bibitem[\protect\citeauthoryear{Gao, Lu \& Zhang}{2010}]{gao} Gao H., Lu Y., Zhang S. N., 2010, ApJ, 717, 268
\bibitem[\protect\citeauthoryear{Gehrels \& Razzaque}{2013}]{gehrels2} Gehrels N., Razzaque S., 2013, Front. Phys., 8, 661
\bibitem[\protect\citeauthoryear{Gehrels et al.}{2006}]{gehrels} Gehrels N. et al., 2006, Nature, 444, 1044
\bibitem[\protect\citeauthoryear{Golkhou \& Butler}{2014}]{golkhou} Golkhou V. Z., Butler N. R., 2014, ApJ, 787, 90
\bibitem[\protect\citeauthoryear{Golkhou, Butler \& Littlejohns}{2015}]{golkhou2} Golkhou V. Z., Butler N. R., Littlejohns O. M., 2015, ApJ, in press (\href{http://arxiv.org/abs/1501.05948}{arXiv:1501.05948})
\bibitem[\protect\citeauthoryear{Gruber}{2012}]{gruber2} Gruber D., 2012, PoS (GRB2012), 007
\bibitem[\protect\citeauthoryear{Gruber et al.}{2014}]{gruber} Gruber D. et al., 2014, ApJS, 211, 12
\bibitem[\protect\citeauthoryear{Hakkila et al.}{2003}]{hakkila} Hakkila J., Giblin T. W., Roiger R. J., Haglin D. J., Paciesas W. S., Meegan C. A., 2003 ApJ, 582, 320
\bibitem[\protect\citeauthoryear{Heussaff, Atteia \& Zolnierowski}{2013}]{heussaff} Heussaff V., Atteia J.-L., Zolnierowski Y., 2013, A\&A, 557, A100
\bibitem[\protect\citeauthoryear{Horv\'ath}{1998}]{horvath98} Horv\'ath I., 1998, ApJ, 508, 757
\bibitem[\protect\citeauthoryear{Horv\'ath}{2002}]{horvath02} Horv\'ath I., 2002, A\&A, 392, 791
\bibitem[\protect\citeauthoryear{Horv\'ath}{2009}]{horvath09} Horv\'ath I., 2009, Ap\&SS, 323, 83
\bibitem[\protect\citeauthoryear{Horv\'ath et al.}{2004}]{horvath04} Horv\'ath I., M\'esz\'aros A., Bal\'azs L. G., Bagoly Z., 2004, Balt. Astron., 13, 217 
\bibitem[\protect\citeauthoryear{Horv\'ath et al.}{2006}]{horvath06} Horv\'ath I., Bal\'azs L. G., Bagoly Z., Ryde F., M\'esz\'aros A., 2006, A\&A, 447, 23
\bibitem[\protect\citeauthoryear{Horv\'ath et al.}{2008a}]{horvath08a} Horv\'ath I., Bal\'azs L. G., Bagoly Z., Veres P., 2008a, A\&A, 489, L1
\bibitem[\protect\citeauthoryear{Horv\'ath et al.}{2008b}]{horvath08b} Horv\'ath I., Bal\'azs L. G., Bagoly Z., Kelemen J., Veres P., Tusn\'ady G., 2008b, AIP Conf. Proc., 1065, 67
\bibitem[\protect\citeauthoryear{Horv\'ath et al.}{2010}]{horvath10} Horv\'ath I., Bagoly Z., Bal\'azs L. G., de Ugarte Postigo A., Veres P., M\'esz\'aros A., 2010, ApJ, 713, 552
\bibitem[\protect\citeauthoryear{Horv\'ath et al.}{2012}]{horvath12} Horv\'ath I., Bal\'azs L. G., Hakkila J., Bagoly Z., Preece R. D., 2012, PoS (GRB2012), 046
\bibitem[\protect\citeauthoryear{Huja \& \v R\'{\i}pa}{2009}]{huja2} Huja D., \v R\'{\i}pa J., 2009, Balt. Astron., 18, 311
\bibitem[\protect\citeauthoryear{Huja, M\'esz\'aros \& \v R\'{\i}pa}{2009}]{huja} Huja D., M\'esz\'aros A., \v R\'{\i}pa J., 2009, A\&A, 504, 67
\bibitem[\protect\citeauthoryear{Hurst}{1951}]{hurst} Hurst, H. E., 1951, Trans. Am. Soc. Civ. Eng., 116, 770
\bibitem[\protect\citeauthoryear{Janiuk et al.}{2006}]{janiuk} Janiuk A., Czerny B., Moderski R., Cline D. B., Matthey C., Otwinowski S., 2006, MNRAS, 365, 874
\bibitem[\protect\citeauthoryear{Kanji}{2006}]{kanji} Kanji G. K., 2006, 100 Statistical Tests. 3rd edn. SAGE Publications
\bibitem[\protect\citeauthoryear{Kendall \& Stuart}{1973}]{kendall} Kendall M., Stuart A., 1973, The Advanced Theory of Statistics. Griffin, London
\bibitem[\protect\citeauthoryear{Klebesadel, Strong \& Olson}{1973}]{klebesadel} Klebesadel R. W., Strong I. B., Olson R. A., 1973, ApJ, 182, L85
\bibitem[\protect\citeauthoryear{Massey}{1951}]{kolmogorov} Massey F. J. Jr, 1951, J. Amer. Statist. Assoc., 46, 68
\bibitem[\protect\citeauthoryear{Koshut et al.}{1996}]{koshut} Koshut T. M., Paciesas W. S., Kouveliotou C., van Paradijs J., Pendleton G. N., Fishman G. J., Meegan C. A., 1996, ApJ, 463, 570
\bibitem[\protect\citeauthoryear{Kouveliotou et al.}{1993}]{kouve} Kouveliotou C., Meegan C. A., Fishman G. J., Bhat N. P., Briggs M. S., Koshut T. M., Paciesas W. S., Pendleton G. N., 1993, Apj, 413, L101
\bibitem[\protect\citeauthoryear{Levene}{1960}]{levene} Levene H., 1960, in Olkin I., Hotelling H., et alia. Contributions to Probability and Statistics, Stanford University Press, p. 278
\bibitem[\protect\citeauthoryear{L\"u et al.}{2010}]{lu} L\"u H.-J., Liang E.-W., Zhang B.-B., Zhang B., 2010, ApJ, 725, 1965
\bibitem[\protect\citeauthoryear{MacLachlan et al.}{2012}]{maclach12} MacLachlan G. A., Shenoy A., Sonbas E., Dhuga K. S., Eskandarian A., Maximon L. C., Parke W. C., 2012, MNRAS, 425, L32
\bibitem[\protect\citeauthoryear{MacLachlan et al.}{2013a}]{maclach13a} MacLachlan G. A. et al., 2013a, MNRAS, 432, 857
\bibitem[\protect\citeauthoryear{MacLachlan et al.}{2013b}]{maclach13b} MacLachlan G. A., Shenoy A., Sonbas E., Coyne, R., Dhuga K. S., Eskandarian A., Maximon L. C., Parke W. C., 2013b, MNRAS, 436, 2907
\bibitem[\protect\citeauthoryear{Magliocchetti, Ghirlanda \& Celotti}{2003}]{maglio} Magliocchetti M., Ghirlanda G., Celotti A., 2003, MNRAS, 343, 255
\bibitem[\protect\citeauthoryear{Mandelbrot \& van Ness}{1968}]{mandel68} Mandelbrot B. B., van Ness J. W., 1968, SIAM Rev., 10, 422
\bibitem[\protect\citeauthoryear{Mandelbrot \& Wallis}{1969}]{mandel69} Mandelbrot, B. B., Wallis, J. R., Water Resour. Res., 4, 909
\bibitem[\protect\citeauthoryear{Mann \& Whitney}{1947}]{mann} Mann H. B., Whitney D. R., 1947, Ann. Math. Stat., 18, 50
\bibitem[\protect\citeauthoryear{Mazets et al.}{1981}]{mazets} Mazets E. P. et al., 1981, Ap\&SS, 80, 3
\bibitem[\protect\citeauthoryear{Meegan et al.}{1992}]{meegan92} Meegan C. A., Fishman G. J., Wilson R. B., Horack J. M., Brock M. N., Paciesas W. S., Pendleton G. N., Kouveliotou C., 1992, Nature, 355, 143
\bibitem[\protect\citeauthoryear{Meegan et al.}{1996}]{meegan96} Meegan C. A. et al., 1996, ApJS, 106, 65
\bibitem[\protect\citeauthoryear{M\'esz\'aros et al.}{2000}]{meszaros4} M\'esz\'aros A., Bagoly Z., Horv\'ath I., Bal\'azs L. G., Vavrek R., 2000, ApJ, 539, 98
\bibitem[\protect\citeauthoryear{M\'esz\'aros, Bagoly \& Vavrek}{2000}]{meszaros3} M\'esz\'aros A., Bagoly Z., Vavrek R., 2000, A\&A, 354, 1
\bibitem[\protect\citeauthoryear{M\'esz\'aros \& \v Sto\v cek}{2003}]{meszaros2} M\'esz\'aros P., \v Sto\v cek J., 2003, A\&A, 403, 443
\bibitem[\protect\citeauthoryear{M\'esz\'aros \& Rees}{2015}]{meszaros} M\'esz\'aros P., Rees M. J., 2015, in Ashtekar A., Berger B., Isenberg J. \& MacCallum M. A. H., eds, General Relativity and Gravitation: A Centennial Perspective, Cambridge University Press (\href{http://arxiv.org/abs/1401.3012}{arXiv:1401.3012})
\bibitem[\protect\citeauthoryear{Mukherjee et al.}{1998}]{mukh} Mukherjee S., Feigelson E. D., Jogesh Babu G., Murtagh F., Fraley C., Raftery A., 1998, ApJ, 508, 314
\bibitem[\protect\citeauthoryear{Muccino et al.}{2013}]{muccino} Muccino M., Ruffini R., Bianco C. L., Izzo L.,  Penacchioni A. V., 2013, ApJ, 763, 125
\bibitem[\protect\citeauthoryear{Nakar}{2007}]{nakar} Nakar E., 2007, Phys. Rep., 442, 166
\bibitem[\protect\citeauthoryear{Paciesas et al.}{1999}]{paciesas} Paciesas W. S. et al., 1999, ApJS, 122, 465
\bibitem[\protect\citeauthoryear{Peng et al.}{1994}]{peng94} Peng, C.-K., Buldyrev, S. V., Havlin, S., Simons, M., Stanley, H. E., Goldberger A. L., 1994, Phys. Rev. E, 49, 1685
\bibitem[\protect\citeauthoryear{Peng et al.}{1995}]{peng95} Peng, C.-K., Havlin, S., Stanley, H., Goldberger, A. L., 1995, Chaos, 5, 82
\bibitem[\protect\citeauthoryear{Qin et al.}{2013}]{qin} Qin Y. et al., 2013, ApJ, 763, 15
\bibitem[\protect\citeauthoryear{Racusin et al.}{2011}]{racusin} Racusin J. L. et al., 2011, ApJ, 738, 138
\bibitem[\protect\citeauthoryear{Press et al.}{2007}]{recipes} Press W. H., Teukolsky S. A., Vetterling W. T., Flannery B. P., 2007, Numerical Recipes. Cambridge University Press
\bibitem[\protect\citeauthoryear{\v R\'{\i}pa et al.}{2009}]{ripa} \v R\'{\i}pa J., M\'esz\'aros A., Wigger C., Huja D., Hudec R., Hajdas W., 2009, A\&A, 498, 399
\bibitem[\protect\citeauthoryear{\v R\'{\i}pa et al.}{2012}]{ripa2} \v R\'{\i}pa J., M\'esz\'aros A., Veres P., Park I. H., 2012, ApJ, 756, 44
\bibitem[\protect\citeauthoryear{Shapiro \& Wilk}{1965}]{shapiro} Shapiro S. S., Wilk M. B., 1965, Biometrika, 52, 591
\bibitem[\protect\citeauthoryear{Simonsen, Hansen \& Nes}{1998}]{simons} Simonsen, I., Hansen, A., Nes O. M., 1998, Phys. Rev. E, 58, 2779
\bibitem[\protect\citeauthoryear{Student}{1908}]{student} Student, 1908, Biometrika, 6, 1
\bibitem[\protect\citeauthoryear{Tarnopolski}{2015a}]{Tarnopolski} Tarnopolski M., 2015a, A\&A, 581, A29
\bibitem[\protect\citeauthoryear{Tarnopolski}{2015b}]{Tarnopolski3} Tarnopolski M., 2015b, preprint (\href{http://arxiv.org/abs/1506.07801}{arXiv:1506.07801})
\bibitem[\protect\citeauthoryear{Tarnopolski}{2015c}]{Tarnopolski2} Tarnopolski M., 2015c, Ap\&SS, 359:20
\bibitem[\protect\citeauthoryear{Woosley \& Bloom}{2006}]{woosley} Woosley S. E., Bloom J. S., 2006, ARA\&A, 44, 507
\bibitem[\protect\citeauthoryear{Vavrek et al.}{2008}]{vavrek} Vavrek R., Bal\'azs L. G., M\'esz\'aros A., Horv\'ath I., Bagoly Z., 2008, MNRAS, 391, 1741
\bibitem[\protect\citeauthoryear{Veres et al.}{2010}]{veres} Veres P., Bagoly Z., Horv\'ath I., M\'esz\'aros A., Bal\'azs G., 2010, ApJ, 725, 1955
\bibitem[\protect\citeauthoryear{Voinov, Nikulin \& Balakrishnan}{2013}]{voinov} Voinov V., Nikulin M., Balakrishnan N., 2013, Chi-Squared Goodness of Fit Tests with Applications. Academic Press
\bibitem[\protect\citeauthoryear{von Kienlin et al.}{2014}]{kienlin} von Kienlin A. et al., 2014, ApJS, 211, 13
\bibitem[\protect\citeauthoryear{Zhang}{2011}]{zhang4} Zhang B., 2011, C. R. Physique, 12, 206
\bibitem[\protect\citeauthoryear{Zhang et al.}{2007}]{zhang} Zhang B., Zhang B.-B., Liang E.-W., Gehrels N., Burrows D. N., M\'esz\'aros P., 2007, ApJ, 655, L25
\bibitem[\protect\citeauthoryear{Zhang et al.}{2009}]{zhang5} Zhang B. et al., 2009, ApJ, 703, 1696
\bibitem[\protect\citeauthoryear{Zhang et al.}{2012}]{zhang3} Zhang F.-W., Shao L., Yan J.-Z., Wei D.-M., 2012, ApJ, 750, 88
\end{thebibliography}
\end{document}